\documentclass[numberedappendix]{emulateapj}
\usepackage{natbib}
\newcommand{\msol}{\,\textrm{M}_\sun}                % Solar mass
\newcommand{\name}{RG1M0150}
\shorttitle{Discovery of a Strongly Lensed Massive Quiescent Galaxy at $z = 2.636$}
\shortauthors{Newman et al.}
\submitted{Accepted to ApJ Letters}
\received{2015 September 16}
\accepted{2015 October 12}
\begin{document}
\title{Discovery of a Strongly Lensed Massive Quiescent Galaxy at $z = 2.636$:\\
Spatially Resolved Spectroscopy and Indications of Rotation}
\author{Andrew B. Newman$^1$, Sirio Belli$^{2,}$\altaffilmark{3}, and Richard S. Ellis$^{2,}$\altaffilmark{4}}
\affil{$^1$ The Observatories of the Carnegie Institution for Science, 813 Santa Barbara St., Pasadena, CA 91101, USA; \href{mailto:anewman@obs.carnegiescience.edu}{anewman@obs.carnegiescience.edu}}  
\affil{$^2 $ Department of Astrophysics, California Institute of Technology, MS 249-17, Pasadena, CA 91125, USA}
\altaffiltext{3}{Present address: Max-Planck-Institut f\"ur extraterrestrische Physik (MPE), Giessenbachstrasse 1, 85748 Garching, Germany}
\altaffiltext{4}{Present address: European Southern Observatory (ESO), Karl-Schwarzschild-Strasse 2, 85748 Garching, Germany}

\begin{abstract}
We report the discovery of \name, a massive, recently quenched galaxy at $z=2.636$ that is multiply imaged by the cluster MACSJ0150.3-1005. We derive a stellar mass of $\log M_*=11.49^{+0.10}_{-0.16}$ and a half-light radius of $R_{e,\rm maj}=1.8\pm0.4$~kpc. Taking advantage of the lensing magnification, we are able to spatially resolve a remarkably massive yet compact quiescent galaxy at $z>2$ in ground-based near-infrared spectroscopic observations using Magellan/FIRE and Keck/MOSFIRE. We find no gradient in the strength of the Balmer absorption lines over $0.6 R_e - 1.6 R_e$, which are consistent with an age of 760~Myr. Gas emission in [\ion{N}{2}] broadly traces the spatial distribution of the stars and is coupled with weak H$\alpha$ emission (log~[\ion{N}{2}]/H$\alpha=0.6\pm0.2$), indicating that OB stars are not the primary ionizing source. The velocity dispersion within the effective radius is $\sigma_{e,\rm stars}=271\pm41$~km~s${}^{-1}$. We detect rotation in the stellar absorption lines for the first time beyond $z\sim1$. Using a two-integral Jeans model that accounts for observational effects, we measure a dynamical mass of $\log M_{\rm dyn}=11.24\pm0.14$ and $V/\sigma=0.70\pm0.21$. This is a high degree of rotation considering the modest observed ellipticity of $0.12\pm0.08$, but it is consistent with predictions from dissipational merger simulations that produce compact remnants. The mass of \name~implies that it is likely to become a slowly rotating elliptical. If it is typical, this suggests that the progenitors of massive ellipticals retain significant net angular momentum after quenching which later declines, perhaps through accretion of satellites.
\end{abstract} 

\section{Introduction}

Observations are beginning to reveal the bulk properties of the high-redshift progenitors of massive elliptical galaxies. Massive quiescent galaxies are present at $z>2$ but have very compact sizes compared to local early-type galaxies (ETGs). Much effort in recent years has focused on understanding the growth of these ``nuggets'' and the emergence of extended stellar envelopes \citep[e.g.,][]{Trujillo06,vanDokkum10,Newman12,Patel13,Belli15}. Recent observations are also beginning to link these early quiescent galaxies to star-forming progenitors at even earlier epochs via their masses, sizes, and kinematics \citep{Barro14,Toft14,vanDokkum15}.

Numerical simulations have made increasingly detailed predictions for the properties of ETGs formed in disk mergers \citep[e.g.,][]{Cox06,Robertson06,Hopkins09,Zolotov15}. Forming a dense stellar core requires significant dissipation, and such gas-rich mergers are predicted to produce rotating remnants---possibly with a velocity exceeding the bulk of today's ETGs \citep{Wuyts10}. Different channels for the formation of compact galaxies can be tested with stellar population gradients \citep{Wellons15} and kinematic data. However, such comparisons require \emph{spatially resolved} observations, which are very challenging given  the small angular sizes of typical $z>2$ quiescent galaxies ($\sim0\farcs1-0\farcs2$). Gravitational lensing provides a promising route toward resolving these compact systems. The main challenge in locating lensed examples is the need for wide-area near-infrared (NIR) imaging to overcome their rarity and faintness at optical wavelengths. Thus far only three example have been published \citep{Muzzin12,Geier13}, and none has spatially resolved spectroscopic data.

In this Letter we present the discovery of \name, a rare example of a massive, quiescent galaxy at $z = 2.636$ that is multiply imaged by the cluster MACSJ0150.3-1005 (R.A.~$1^{\rm h}50^{\rm m}21^{\rm s}.3$, Decl.~$-10^{\circ}05'30''$, $z=0.365$; \citealt{Ebeling01}). \name~was located in a search for bright sources with red NIR colors that are magnified by clusters in the \emph{Hubble Space Telescope} (\emph{HST}) archive. It is extremely bright ($K_{s,\rm AB}=19.2$) and well suited for detailed follow-up observations. We analyze NIR spectra collected with Magellan/FIRE and Keck/MOSFIRE and are able to spatially resolve stellar absorption lines with spectral resolution sufficient for kinematic work. We use these unique data to study the resolved stellar populations and rotational support of \name.

Throughout we adopt AB magnitudes and the cosmological parameters $(\Omega_m,\Omega_v,h)=(0.3,0.7,0.7)$.

\begin{figure*}
\centering
\includegraphics[width=\linewidth]{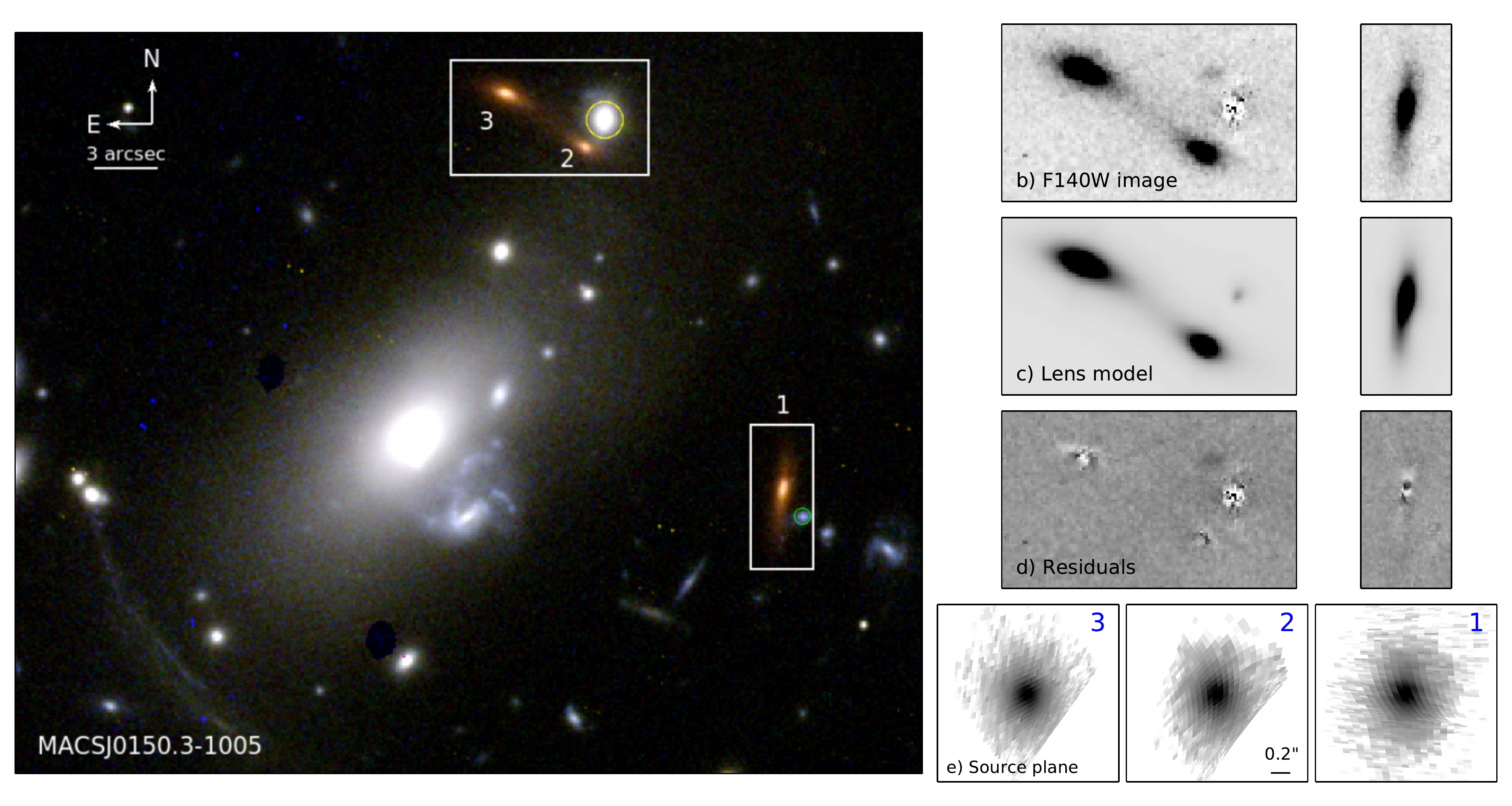}
\caption{\emph{Left:} F814W/F110W/F140W composite image of MACSJ0150.3-1005 with an arcsinh stretch. Images 1, 2, and 3 of \name~are marked. \emph{Right:} Zooms on the observed F140W images (panel b, linear stretch) are compared to the best-fitting S\'{e}rsic model that has been lensed and PSF-convolved (panel c) and the residuals (panel $\rm d = b - c$). Panel e (arcsinh stretch) shows the PSF-deconvolved images in the source plane, constructed by adding the unconvolved image plane model to the residuals (i.e., ${\rm Data} - {{\rm Model}_{\rm convolved}} + {\rm Model_{\rm unconvolved}}$) and casting pixels to the source plane.\label{fig:finder}}
\end{figure*}

\pagebreak
\section{Observations}

\subsection{Imaging\label{sec:imaging}}

\name~was identified in a search of archival \emph{HST}/WFC3 images of massive clusters. We targeted bright, magnified sources with red NIR colors indicative of a Balmer/4000~\AA~break at $z\gtrsim2$. \emph{HST} imaging of \name~covers WFC3/F140W, WFC3/F110W, ACS/F814W, and WFCP2/F606W (SNAP-12884 and SNAP-11103, P.I.~Ebeling; see Figure~\ref{fig:finder}). We also obtained $J$, $H$, and $K_s$ band observations with Magellan/FourStar \citep{Persson13} which were reduced following procedures discussed by \citet{Kelson14}. Colors were measured in a $0\farcs6\times0\farcs75$ rectangular aperture along Image~1 matching the spectroscopic aperture used in Section~\ref{sec:specdata}. Matching of the point spread function (PSF) between the FourStar and \emph{HST} images was performed as described by \cite{Newman12}. Total fluxes were obtained by scaling to the F140W flux in larger $8'' \times 2\farcs5$ rectangle. We also incorporate 3.4$\mu$m and 4.6$\mu$m measurements from the AllWISE catalog.\footnote{\url{http://wise2.ipac.caltech.edu/docs/release/allwise/}} Zeropoint and PSF matching uncertainties of 0.03~mag (\emph{HST}), 0.05~mag (FourStar), and 0.1~mag (\emph{WISE}) were added in quadrature.

\subsection{Spectroscopy\label{sec:specdata}}

We obtained NIR spectroscopic observations of Image~1 of \name~with both MOSFIRE \citep{McLean12} at the Keck~I telescope and FIRE \citep{Simcoe13} at the Magellan Baade telescope. FIRE is a cross-dispersed echellette instrument that covers $0.8-2.45~\mu$m simultaneously. The MOSFIRE data cover the $H$ band only but afford somewhat cleaner sky subtraction owing to the faster dithering cadence. A position angle (PA) of $-7.5^{\circ}$ was selected to place the slit along Image~1.

The MOSFIRE exposures total 260~minutes taken over 2014 November~26-27, consisting of 2~minute integrations in an AB dither pattern. The $0\farcs7$ slit provided a resolution of $\sigma_{\rm inst}=36~{\rm km}~{\rm s}^{-1}$. Another slit was positioned on a star to measure the average seeing of $0\farcs67$. Data were reduced using the MOSFIRE Data Reduction Pipeline.
 
The FIRE observations were conducted over 2014 November~1 and 3 and total 390~minutes, consisting of 30~minute exposures using up-the-ramp sampling. The $6''\times0\farcs75$ slit provided a resolution of $\sigma_{\rm inst} = 30~{\rm km}~{\rm s}^{-1}$. The average seeing of $0\farcs55$ was determined by monitoring a star on the NIR slit viewing camera. The data were reduced using a modified version of the FIREHOSE pipeline that produces two-dimensional rectified spectra. For both data sets, telluric absorption correction and flux calibration were performed using A0V star observations and {\tt xtellcor} \citep{Vacca03}. %Sky subtraction was performed by modeling the two-dimensional sky and object spectra with b-spline techniques. The spatial profile of Image~1 in each exposure was estimated by shifting and convolving the WFC3/F140W profile.

\begin{figure*}
\centering
\includegraphics[width=0.4\linewidth]{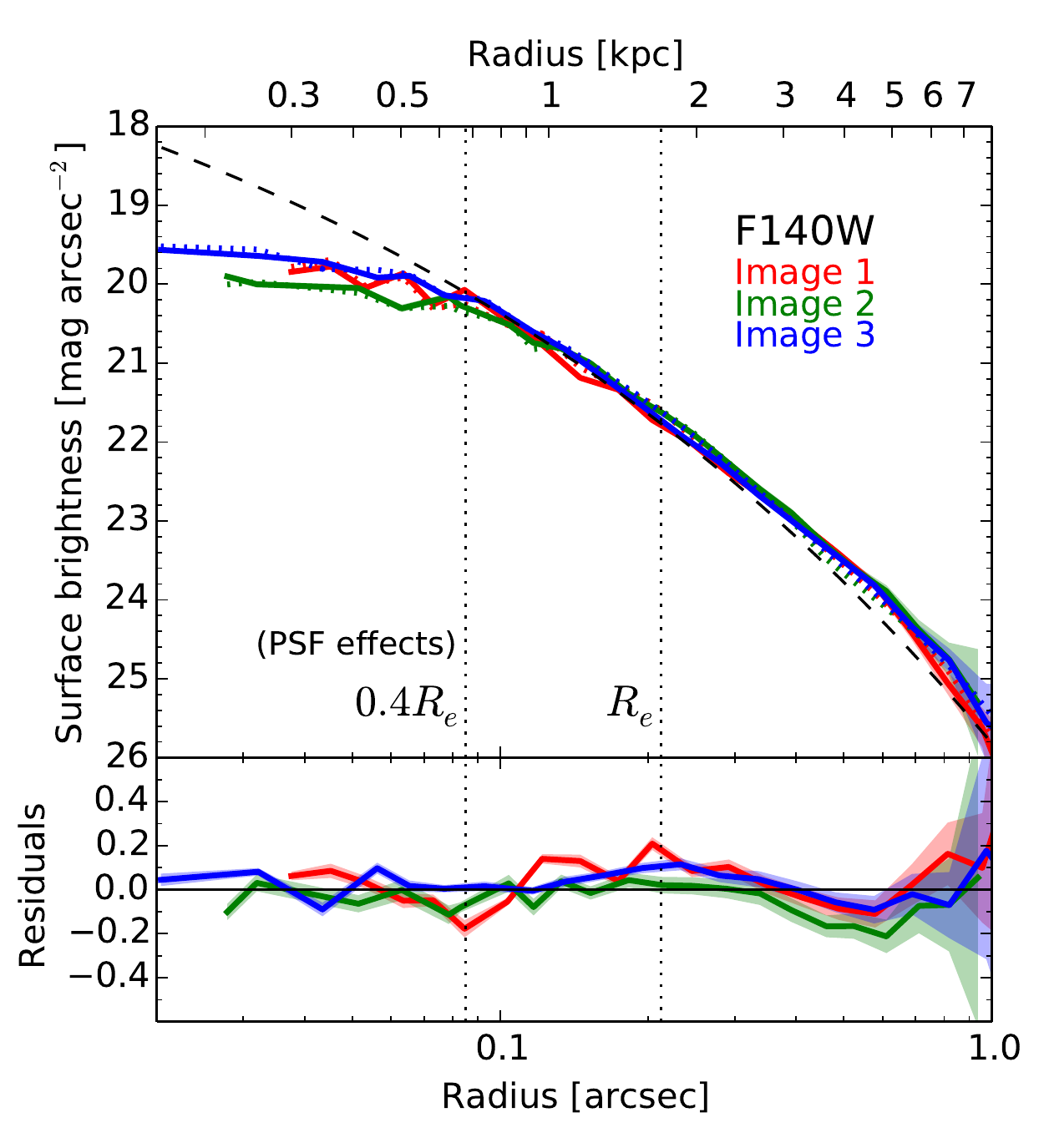}
\includegraphics[width=0.4\linewidth]{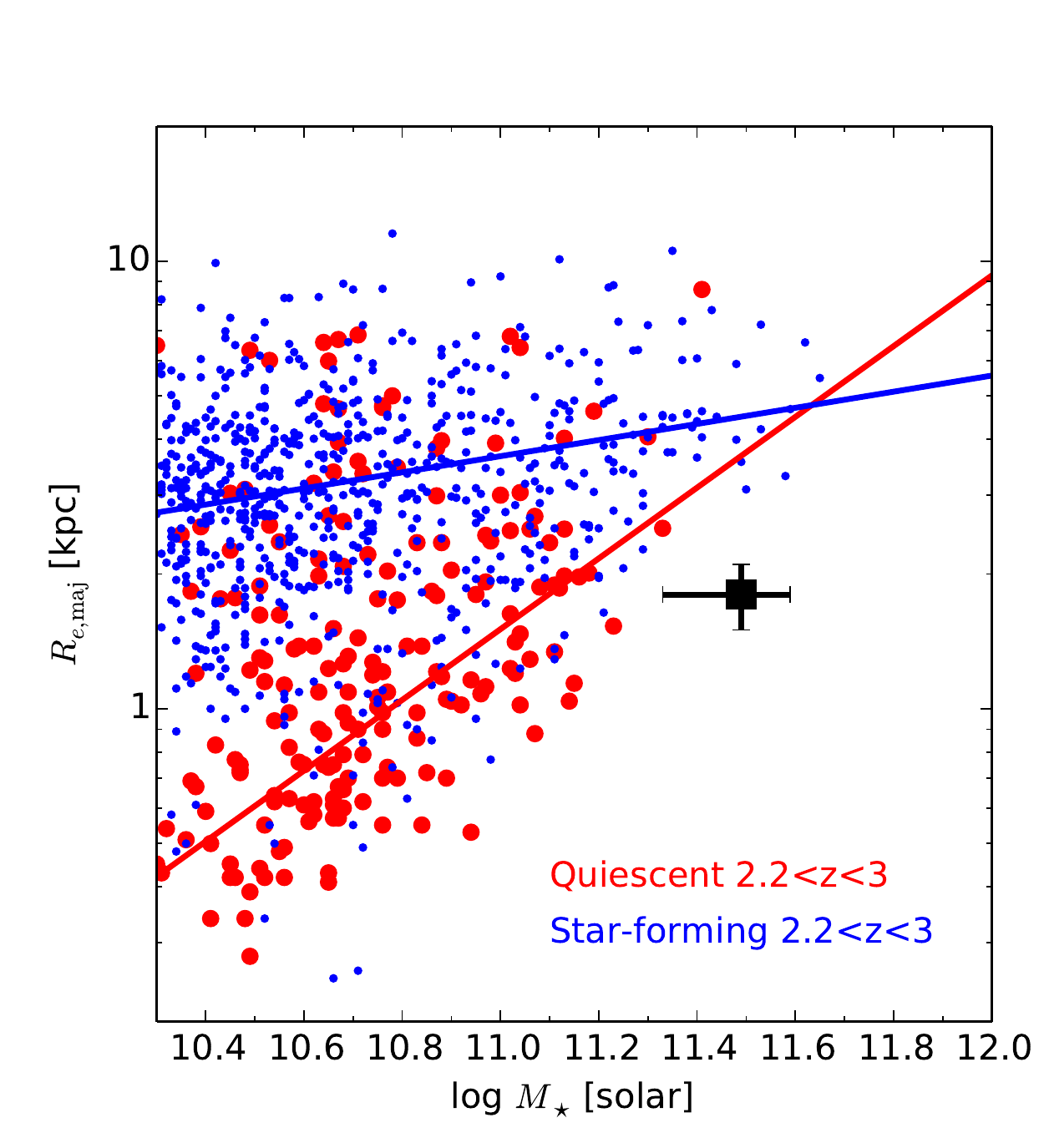}
\caption{\emph{Left:} Source plane F140W surface brightness profiles reconstructed from each image of \name~(solid colored lines). Dotted colored lines show the corresponding reconstructions from the PSF-convolved S\'{e}rsic model fits, with residuals indicated in the lower panel. (At small radii, the reconstructions differ because the PSF affects each image differently.) Shading indicates the $1\sigma$ uncertainty. The black dashed line shows the unconvolved S\'{e}rsic model. \emph{Right:} The compact size of \name~(black) is compared to $2.2<z<3$ field galaxies drawn from the 3D-HST survey catalogs \citep{Skelton14} with structural parameters from \citet{vanderWel14}.\label{fig:sbprofile}}
\end{figure*}

\subsection{Lens Model}

We construct a lens model using the three images of \name~as constraints. Analytic models  describe the lens mass distribution and the source light profile. Their parameters are constrained simultaneously using the ray-tracing code described in \citet{Newman15}, which fits the pixel-level WFC3/F140W data. 

Given its smooth, symmetric appearance, the source was modeled using an elliptical S\'{e}rsic profile. To model the cluster lens, we use several dual pseudo-isothermal elliptical (dPIE) mass distributions (see \citealt{Eliasdottir07}). Each is described by two scale radii $r_{\rm cut}$ and $r_{\rm core}$ and a normalization $\sigma_0$, in addition to its center, PA, and ellipticity. All parameters of the dPIE describing the cluster dark matter halo are free other than $r_{\rm cut}$, which is beyond the strong lensing zone and can be fixed to 1~Mpc \citep[e.g.,][]{Richard10}. Additional dPIE components account for deflection by luminous cluster galaxies and are generally modeled using the scaling relations $\sigma_0=\sigma_{0}^{\star}(L/L_*)^{1/4}$ and $r_{\rm cut}=r_{\rm cut}^{\star}(L/L_*)^{1/2}$, with priors on $\sigma_0^{\star}$ and $r_{\rm cut}^{\star}$ as described by \citet{Newman13}. However, the brightest cluster galaxy and the two galaxies circled in Figure~\ref{fig:finder}, which are important perturbers, were allowed to depart from these scaling relations. We also included external shear to fine tune the image reconstruction, although this did not make a major difference to our final model.
 
Figure~\ref{fig:finder} demonstrates that our best-fitting model produces three images that match the F140W data well. The model produces areal magnifications of $\mu_1=3.9^{+1.6}_{-0.7}$, $\mu_2=2.4^{+1.5}_{-0.5}$ and $\mu_3=4.4^{+2.0}_{-0.8}$ for Images 1, 2, and 3, respectively. Uncertainties were conservatively estimated by comparing to an independent set of {\tt Lenstool} models \citep{Kneib93,Jullo07} constrained only by the positions, ellipticity, and fluxes of the images and varying the mass model components.

\section{Structure and Stellar Population}

\subsection{Stellar Structure}

Our best-fitting S\'{e}rsic model indicates that \name~is a compact, nearly round, concentrated galaxy with an effective radius $R_{e, \rm maj}=0\farcs23\pm0\farcs05=1.8\pm0.4~{\rm kpc}$ (semimajor axis), S\'{e}rsic index $n=3.5\pm0.9$, magnitude $m_{\rm F140W}=21.8^{+0.4}_{-0.2}$, axis ratio $b/a=0.88\pm0.06$, and ${\rm PA}=2^{\circ}\pm14^{\circ}$. Uncertainties were estimated from the scatter in parameters obtained when fitting the three multiple images individually ($m_{\rm F140W}$ also includes the systematic magnification uncertainty).

We reconstruct source plane surface brightness profiles for each image in Figure~\ref{fig:sbprofile}. The lensing magnification enables us to probe down to scales of $R=700~{\rm pc}=0.4R_e$ before PSF blurring becomes significant. Over the range $R=0.4R_e-4R_e$ a single S\'{e}rsic profile remains adequate to fit the F140W light with systematic residuals of $\lesssim 10\%$ (lower panel). The rest-frame colors of \name~($U-V=1.69\pm0.06$, $V-J=0.73\pm0.14$) satisfy the criterion often used to select quiescent galaxies. The right panel of Figure~\ref{fig:sbprofile} demonstrates that \name~is a compact galaxy, lying slightly below the mass--size relation defined by other quiescent systems at similar redshift.

\begin{figure*}
\centering
\includegraphics[width=\linewidth]{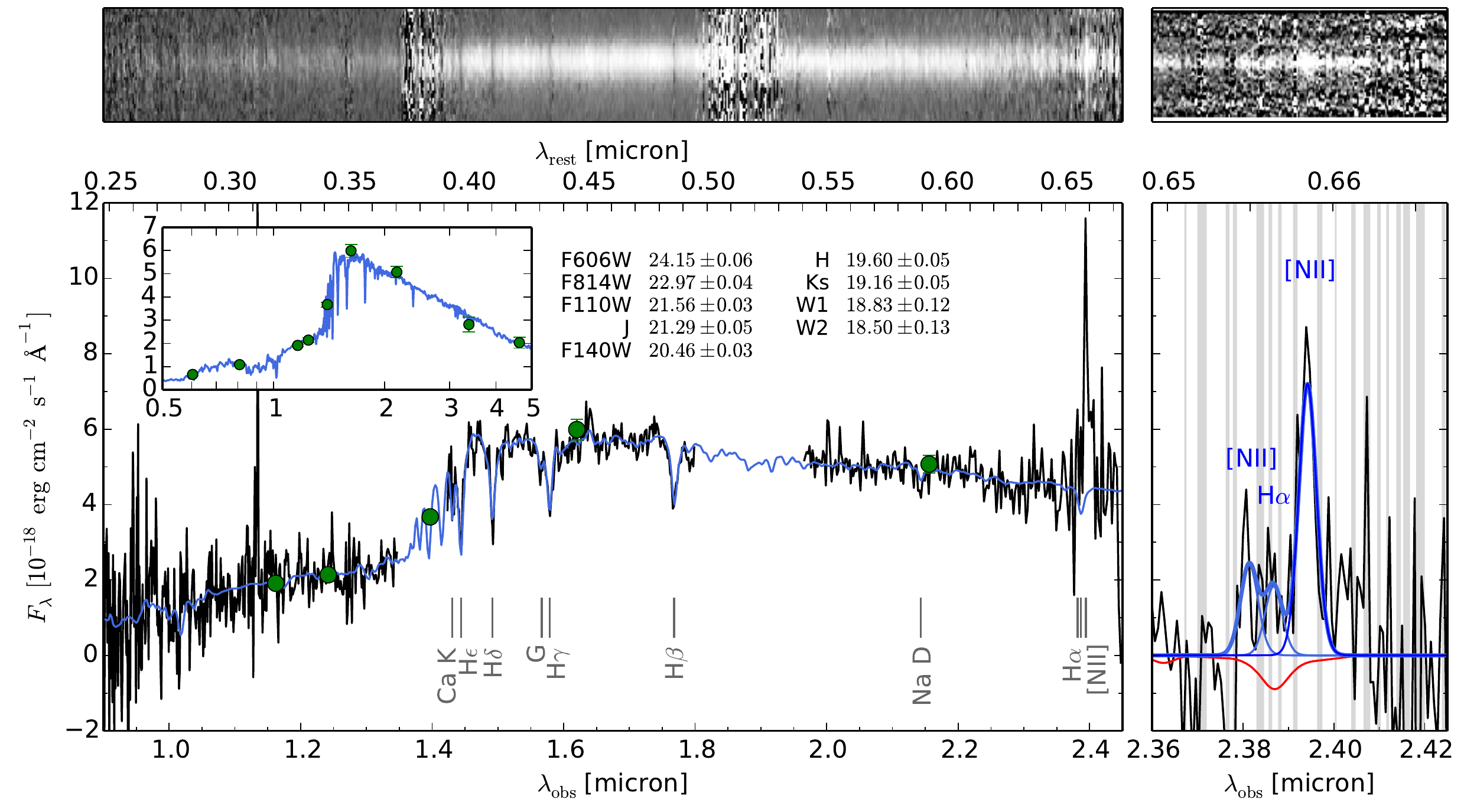}
\caption{\emph{Left:} The FIRE spectrum of Image~1, resampled to 300~km~s${}^{-1}$ bins for clarity, is compared to the stellar population model described in Section~\ref{sec:stellarpop}. Green points show the broadband photometry, which is displayed on an expanded scale in the inset. Resampled two-dimensional spectra are shown above. \emph{Right:} Zoom on the H$\alpha$+[\ion{N}{2}] region with the stellar model (red) subtracted. Blue lines show a Gaussian decomposition with [\ion{N}{2}]$\lambda6585$/[\ion{N}{2}]$\lambda$6550 fixed to 3. The observed spectrum has been scaled by a factor of 3.6 to match the total flux of Image~1. Gray bands denote strong sky lines.\label{fig:sedfit}}
\end{figure*}

\subsection{Stellar Population}
\label{sec:stellarpop}

\name~presents a red continuum shape with strong Balmer absorption dominating the rest-optical spectrum. In Figure~\ref{fig:sedfit} we combine the broadband photometry with our FIRE spectrum to constrain the stellar population using our {\tt pyspecfit} code \citep{Newman14}. We adopt solar metallicity \citet{Bruzual03} models and a \citet{Chabrier03} initial mass function. The FIRE spectrum is extracted in a rectangular aperture extending $\pm0\farcs6$ from the center of Image~1, which best approximates $\pm R_e$ in the source plane. %Photometry in the same aperture was described in Section~\ref{sec:imaging}.

We infer a remarkably high stellar mass of $\log M_*=11.49^{+0.10}_{-0.16}$, scaled to the source plane S\'{e}rsic model flux and including the magnification uncertainty. Given the very high indicated mass, we simulate the probability distribution of the most massive $2 < z < 3$ quiescent galaxy lensed by a cluster in the \emph{HST}/WFC3 archive (80 clusters, assuming a typical cross-section $\sigma = 200$~arcsec${}^2$ for magnification $\mu>3$ similar to MACSJ0150.3-1005) using the \citet{Muzzin13} mass functions. A lensed galaxy with $\log M_*>11.49$ is indeed quite rare ($p=0.005$). The probability increases to $p=0.02$ if the threshold is lowered by $1\sigma$ to $\log M_*>11.33$. If we take our simulated probability distribution as a prior in order to account for an Eddington-type bias, the posterior for the stellar mass would shift downward to $\log M_*=11.33^{+0.11}_{-0.15}$.

For an exponentially declining star formation history, we find an age of $760 \pm 50$~Myr and a prompt $e$-folding time of $97\pm11$~Myr. The corresponding ongoing star formation rate is only ${\rm SFR}=2\pm1~{\rm M}_{\odot}$~yr${}^{-1}$ (accounting for magnification), indicating that star formation was recently and quickly truncated in this galaxy. The specific star formation rate is $\log {\rm SSFR}/{\rm yr}^{-1}<-11.1$ (68\% CL) independent of magnification. Dust attenuation is moderate with $A_V=0.52 \pm 0.08$ assuming the \citet{Calzetti00} reddening curve. (These uncertainties are purely statistical and do not include known differences among stellar population models.)

%In order to test whether \name~might be experiencing a recent rejuvenation of star formation, we also fit a two-component SFH: a old burst at $z_{\rm form}=8$ combined with a later exponentially-declining SFH. An equally good fit is obtained when the younger population formed $530\pm30$~Myr ago and comprises $44\pm8\%$ of the mass. Thus, even allowing for an underlying old population, a major fraction of the stars in this galaxy seem to have formed in an event 0.5-0.8~Gyr ago, after which star formation decayed rapidly to the observed low level.

\begin{figure*}
\centering
\includegraphics[width=0.48\linewidth]{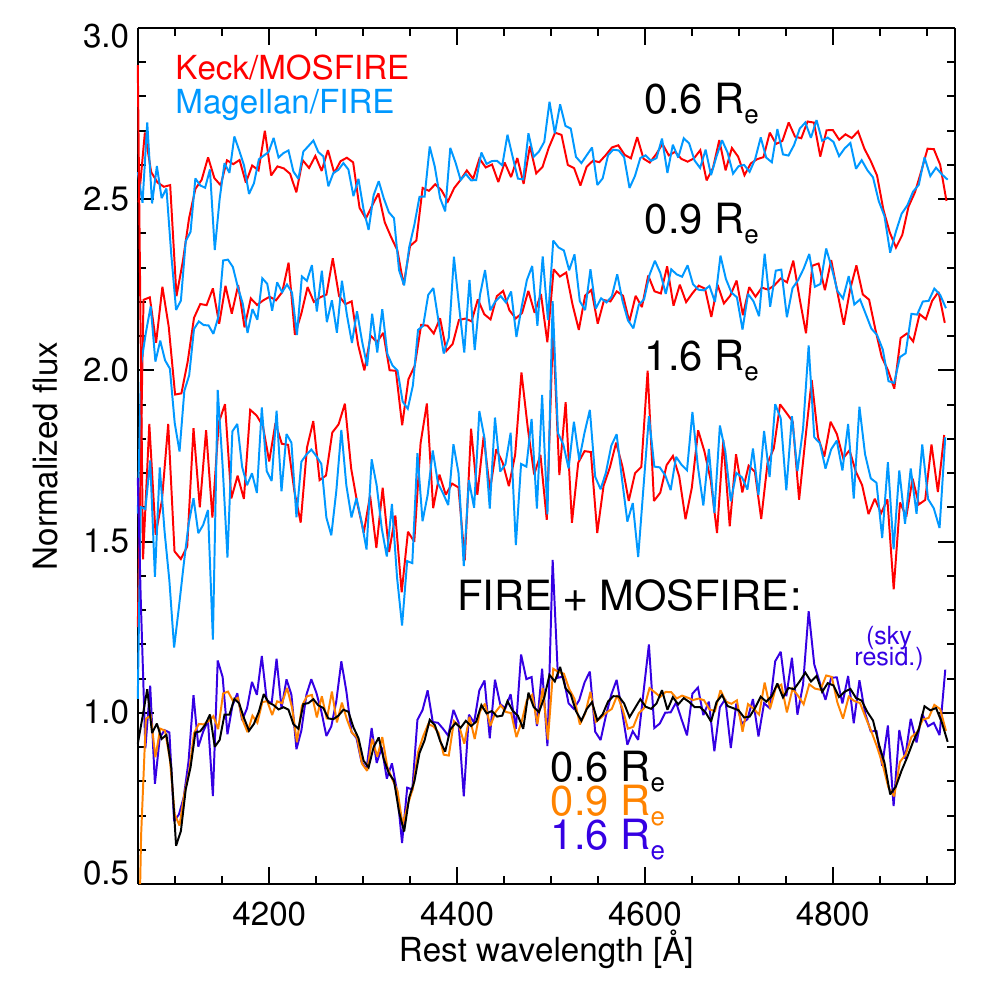}
\includegraphics[width=0.48\linewidth]{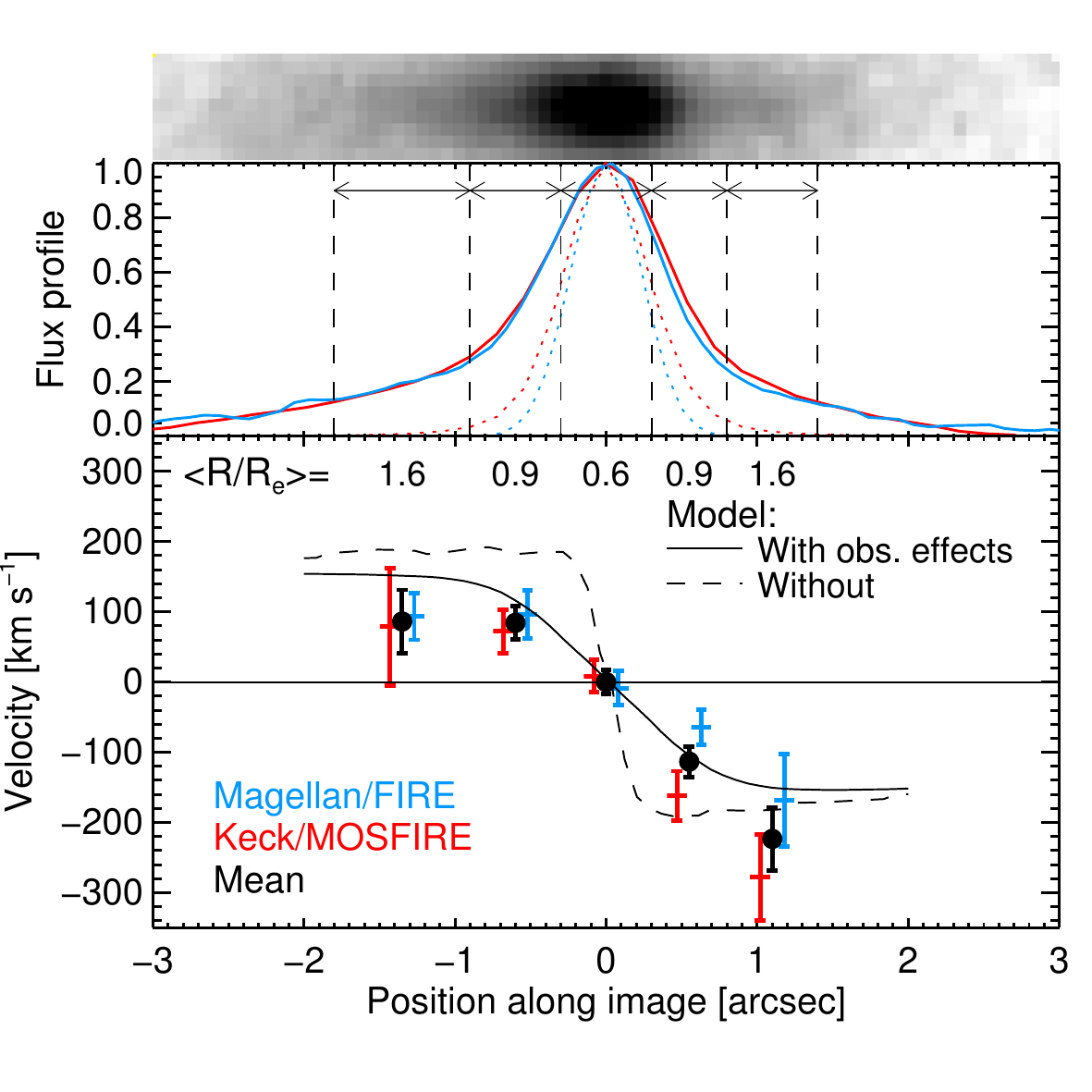}
\caption{\emph{Left:} Resolved Balmer absorption lines extracted in several spatial bins. Corresponding spectra on opposite side of the galaxy center have been shifted to the systemic redshift and coadded. The FIRE and MOSFIRE spectra are combined in the lower spectrum and demonstrate no significant gradients in Balmer line strength over $0.6R_e-1.6R_e$. \emph{Right:} The top panel illustrates the position of the spectrograph slit as seen in the WFC3/F140W image. The middle panel compares the flux distributions in the spectra to the PSFs (dotted), demonstrating that we spatially resolve Image~1. Five extraction bins are indicated. The lower panel indicates the presence of rotation measured using the stellar absorption lines.
\label{fig:stacks}}
\end{figure*}

The lensing magnification presents a unique opportunity to probe the location of the last star formation event by spatially resolving the Balmer absorption. We extract spectra in 5 rectangular apertures whose boundaries along the spectroscopic slit are chosen to examine the same luminosity-weighted radius in the source plane on either side of the galaxy: $\langle R\rangle = 0.6R_e$, $0.9R_e$, and $1.6R_e$. Figure~\ref{fig:stacks} demonstrates that the FIRE and MOSFIRE spectra agree well, and we therefore combine them in the bottom of the panel to search for gradients in Balmer line strength. No variation is apparent, even though the spectra would be sensitive to an age difference of $\simeq200$~Myr. Thus, the last star formation event in \name~was not highly centrally concentrated.

\subsection{Ionized Gas}

We detect emission in the [\ion{N}{2}] $\lambda\lambda6550,6585$ and H$\alpha$ lines (see Figure~\ref{fig:sedfit}). Weak H$\alpha$ emission with a modest equivalent width of ${\rm EW}_{\rm H\alpha}=6\pm 2$~\AA~(rest frame) is evident only after subtracting the underlying stellar absorption. [\ion{N}{2}] is much stronger (${\rm EW}_{[\rm N{\sc II}]\lambda6585}=22\pm2$~\AA). The very high ratio log([\ion{N}{2}]$\lambda 6585 / {\rm H}\alpha)=0.6\pm0.2$ exceeds that attainable through photoionization by a starburst \citep[e.g.,][]{Kewley01}. The main ionizing source is therefore not OB stars, a conclusion consistent with the low star formation rate inferred from the stellar continuum. Although other bright emission line diagnostics are not available (e.g., [\ion{O}{2}] and [\ion{O}{3}] are inaccessible due to atmospheric opacity), similarly high [\ion{N}{2}]/H$\alpha$ ratios are sometimes seen in red sequence and post-starburst galaxies at lower redshift that usually have LINER-type spectra \citep[e.g.,][]{Yan06,Lemaux10}. The [\ion{N}{2}]$\lambda6585$ emission in \name~has a roughly constant EW across the galaxy, which suggests (but does not prove) that the main ionizing source is distributed similarly to the stars (e.g., post-AGB stars or shocks) rather than an active galactic nucleus (AGN). X-ray, radio, and mid/far-IR observations could clarify this by testing for the presence of an AGN or obscured star formation.

\section{Stellar Kinematics: Rotation in a Quiescent $z>2$ Galaxy\label{sec:kinematics}}

Stellar kinematics were measured using the $H$-band ($\lambda_{\rm rest}=4060~{\rm\AA}-4930~{\rm\AA}$) spectra within the spatial apertures described in Section~\ref{sec:stellarpop}. We use the {\tt ppxf} code \citep{Cappellari04} and allow a linear combination of \citet{Bruzual03} simple stellar populations as the template.

\subsection{Velocity Dispersion}

We first determine the velocity dispersion $\sigma_e$ within the aperture encompassing $R_e$ (see Section~\ref{sec:stellarpop}). This is simply the second moment of the line-of-sight velocity distribution and includes both random and streaming motions. The FIRE and MOSFIRE spectra yield consistent estimates of $\sigma_{e,\rm stars}=288\pm27$~km~s${}^{-1}$ and $254\pm25$~km~s${}^{-1}$, respectively, which we average to obtain $\sigma_{e,\rm stars} = 271\pm18$~km~s${}^{-1}$. Tests of several systematic uncertainties revealed that sensitivity to template construction and the inclusion or exclusion of Balmer lines dominate the error budget, contributing $\pm 38$~km~s${}^{-1}$. Combining the uncertainties in quadrature yields $\sigma_{e,\rm stars}=271\pm42$~km~s${}^{-1}$, which is comparable to that measured for the ionized gas in the same aperture, $\sigma_{e,\rm gas}=255\pm32$~km~s${}^{-1}$.

\subsection{Rotation and Dynamical Model}

Although our spectra are insufficient for resolved measurements of $\sigma$, we do have sufficient signal to measure velocities in our five spatial bins. This the first such measurement derived from stellar absorption features at redshifts beyond the $z\sim1$ study of \citet{vanderWel08}. Figure~\ref{fig:stacks} (right panel) shows that all 4 independent measurements on the southern half of the arc are redshifted, while all 4 measurements on the northern half are blueshifted with respect to $z_{\rm sys}=2.6356\pm0.0002$ measured in the central spatial bin. The presence of rotation with the same directional sense in independent data sets gives us significantly higher confidence in its reality. 

To compare the rotational velocity of \name~with other samples, we must account for observational effects, particularly the seeing, slit width, and binning. We do this by constructing a two-integral (semi-isotropic) oblate Jeans dynamical model. We assume mass follows the light distribution described by our S\'{e}rsic model. The free parameters are $M_{\rm dyn}$, the rotation parameter $k$ \citep{Satoh80}, and the inclination. (See \citealt{vanderWel08} for a quite similar approach.) The dynamical model is constructed in the source plane, then lensed, convolved by the seeing, and binned like the data. The model is constrained by the rotation points (Figure~\ref{fig:stacks}) and $\sigma_{e,\rm stars}$. Fortuitously, the slit PA of $-7.5^{\circ}$, which is aligned with the direction of magnification, is close to the major axis PA ($2^{\circ}\pm14^{\circ}$).

The inferred dynamical mass is $\log M_{\rm dyn}=11.24\pm0.14$ when marginalizing over the unknown inclination $0<\cos i<\cos i_{\rm min}$. Here we choose $i_{\rm min}=29^{\circ}$ because it corresponds to a maximum intrinsic ellipticity $e_{\rm int}=0.8$. Although the intrinsic $(V/\sigma)_{\rm int}$ is sensitive to inclination, the projected value that we measure is not. The inferred $M_{\rm dyn}$ is also insensitive to inclination except when the galaxy is close to face-on (this result is generic in Jeans dynamical modeling, e.g., \citealt{vanderMarel91}). In particular, when fixing $i = i_{\rm min}$ we find that that the inferred $M_{\rm dyn}$ would increase significantly by 0.16~dex. Adopting any inclination $i>32^{\circ}$ that is even slightly larger, however, affects $M_{\rm dyn}$ by $<0.05$~dex.

Adopting a local virial calibration $\log M_{\rm vir}=\log5\sigma_e^2 R_e/G=11.19\pm0.12$ would have given a slightly smaller mass. $M_{\rm dyn}$ is smaller than, but consistent with, our estimate of $M_*$. However, some tension would arise had we adopted a Salpeter-type initial mass function. It will be interesting to evaluate this further with larger samples of lensed galaxies.

Observational effects modify the observed rotation significantly, as Figure~\ref{fig:stacks} demonstrates. Our modeling allows us to account for these. The best-fitting model has a maximum projected rotation velocity $V_{\rm max}=189\pm34$~km~s${}^{-1}$ and $V/\sigma=0.70\pm0.21$. This model adopts the classic definition of $V/\sigma$ based on $V_{\rm max}$ and $\sigma_0$ averaged within $R_e/2$ \citep[e.g.,][]{Davies83}.

\begin{figure}
\centering
\includegraphics[width=\linewidth]{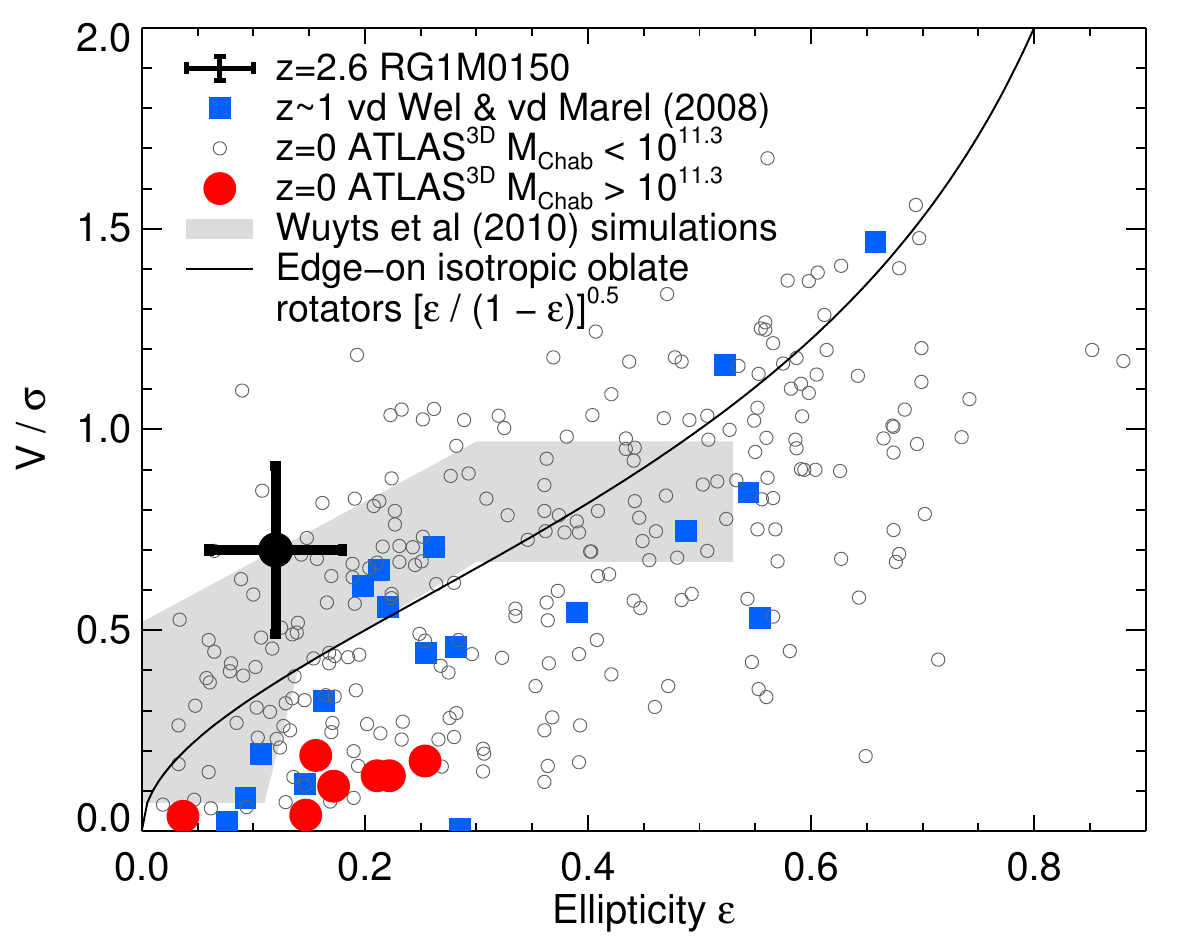}
\caption{Rotational support in \name~is compared to observations of ETGs at $z \sim 1$ \citep{vanderWel08} and $z \sim 0$ \citep[][ATLAS${}^{\rm 3D}$]{Emsellem11}. The solid line is the \citet{Binney78} curve for isotropic oblate rotators. The gray region approximately encompasses the simulated high-$z$ merger remnants from \citet{Wuyts10}. \label{fig:vsigma}}
\end{figure}

\section{Discussion}

A central question in the evolution of ETGs is the extent to which the quenching of star formation and the transformation of stellar structure, morphology, and kinematics proceed together (e.g., in one major merger episode) or in several distinct events. The presence of compact quiescent galaxies at high redshift demonstrates that growth of the stellar distribution proceeds long after quenching. Recent examinations of the evolving ellipticities of quiescent galaxies suggest that these early, compact systems were also more disk-like \citep{vanderWel11,Chang13}. The key test is a direct measurement of rotation, presented in this Letter for the first time.

Figure~\ref{fig:vsigma} compares the rotational support of \name~with that of local ETGs in the classic $(\epsilon,V/\sigma)$ diagram.\footnote{Using the ATLAS${}^{\rm3D}$ kinematic maps, we determined that their adopted definition $(V/\sigma)_e$ \citep[Equation~10 of][]{Cappellari07} is lower than ours by an average factor of 1.53. We apply this correction to place the data in Figure~\ref{fig:vsigma} on a uniform scale.}  \name~has a high $V/\sigma$ for its modest observed ellipticity that places it on the upper envelope of local ETGs. Although we cannot determine the inclination and deproject $V/\sigma$ with current data, it is possible that the edge-on rotation velocity is significantly higher.

Numerical simulations of galaxy mergers require high gas fractions of $\sim40\%$ \citep{Wuyts10} to produce compact remnants such as \name. These simulations generally find that a significant amount of angular momentum is retained and suggest that rotation will be more significant in quiescent galaxies at higher redshift, when mergers were more dissipative. \citet{Wuyts10} predicted that quiescent $z\sim2$ galaxies would fall on the upper envelope of local ETGs in the $(\epsilon,V/\sigma)$ diagram. Intriguingly, we find that \name~indeed lies close in Figure~5 to their simulated merger remnants, which are compact systems with $M_*\sim10^{11}\msol$ and $R_e\sim1$~kpc comparable to typical $z\simeq2$ quiescent galaxies. 

Simulations by \citet{Zolotov15} predict that the gas in compact quiescent galaxies and their immediate star-forming progenitors is kinematically colder than the stars, i.e., has higher $V/\sigma$. This causes the projected $\sigma_{e,\rm stars}/\sigma_{e,\rm gas}$ to be $\sim1$ for an edge-on view but progressively higher for lower inclinations. Supporting this, \citet{Barro15} measured $\sigma_{\rm stars}/\sigma_{\rm gas}=1.7\pm0.5$ in a quenching compact galaxy at $z=1.7$. We find that the stellar and gas line widths are closer in \name, with $\sigma_{e,\rm stars}/\sigma_{e,\rm gas}=1.1\pm0.2$, but larger samples will be needed to test this hypothesis due to the strong inclination dependence.

Nearly all local ETGs with $M_*>10^{11.3}\msol$ (scaled to a Chabrier IMF) are ``slow rotators'' with $V/\sigma$ much smaller than \name~(red points in Figure~\ref{fig:vsigma}). Our best estimate for the stellar mass of \name~already places it within this range. Even allowing a somewhat smaller mass (e.g., our $M_{\rm dyn}$), it is quite likely that \name~will evolve into a slow rotator: empirical estimates based on number density arguments \citep[e.g.,][]{Muzzin13,Patel13} imply that \name~will grow in stellar mass by 0.2-0.4~dex by $z=0$. Stellar populations in such ultra-massive galaxies today are almost exclusively very old \citep[e.g.,][]{McDermid15}, so significant later star formation is unlikely. If \name~is typical, this suggests that the progenitors of giant ellipticals are still rotating significantly after they have been quenched and must reduce their net angular momentum later. A sequence of mostly dry mergers, suspected of driving the observed $\simeq4\times$ growth in effective radius, might provide a physical explanation.

General conclusions about galaxy formation obviously cannot be drawn from \name~alone. Fortunately, a larger sample of lensed compact galaxies at $z\sim2$ (mostly at lower masses) is within reach and provides the most feasible route toward resolving the stellar populations and dynamics of these intriguing objects.

\acknowledgements
We thank Dan Kelson for expertise reducing the FourStar data and insightful discussions. It is a pleasure to acknowledge helpful conversations with Adam Muzzin and Sune Toft, as well as the referee for a prompt and thoughtful report. The authors acknowledge the very significant cultural role that the summit of Mauna Kea has always had within the indigenous Hawaiian community. We are most fortunate to have the opportunity to conduct observations from this mountain.

\end{document}